\documentclass[]{spie}  

\usepackage{amsmath,amsfonts,amssymb}
\usepackage{graphicx}
\usepackage[colorlinks=true, allcolors=blue]{hyperref}

\usepackage{journals}

\title{CUBESPEC: Low-cost space-based astronomical spectroscopy}

\author[a]{Gert Raskin}
\author[b]{Tjorven Delabie}
\author[b]{Wim De Munter}
\author[a]{Hugues Sana}
\author[a]{Bart Vandenbussche}
\author[a]{Bram Vandoren}
\author[c]{Victoria Antoci}
\author[c]{Hans Kjeldsen}
\author[c, d]{Christoffer Karoff}
\author[e]{Alex de Koter}
\author[e]{Jean-Michel D\'{e}sert}
\author[a, f]{Tom Mladenov}
\author[b]{Dirk Vandepitte}

\affil[a]{KU Leuven, Institute of Astronomy, Celestijnenlaan 200D, 3001 Leuven, Belgium}
\affil[b]{KU Leuven,  Department of Mechanical Engineering, Celestijnenlaan 300, 3001 Leuven, Belgium}
\affil[c]{Stellar Astrophysics Centre, Aarhus University, Ny Munkegade 120, DK-8000 Aarhus C, Denmark}
\affil[d]{Department of Geoscience, Aarhus University, H{\o}egh-Guldbergs Gade 2, DK-8000, Aarhus C, Denmark}
\affil[e]{Anton Pannekoek Institute for Astronomy, University of Amsterdam, Science Park 904, 1098 XH Amsterdam, The Netherlands}
\affil[f]{University Hasselt, Faculty of Engineering Technology, 3590 Diepenbeek, Belgium}

\authorinfo{Send correspondence to gert.raskin@kuleuven.be}

\begin{document} 
\maketitle

\begin{abstract}
CubeSats are routinely used for low-cost photometry from space. Space-borne spectroscopy, however, is still the exclusive domain of much larger platforms. Key astrophysical questions in e.g.\ stellar physics and exoplanet research require uninterrupted spectral monitoring from space over weeks or months. Such monitoring of individual sources is unfortunately not affordable with these large platforms. With CUBESPEC we plan to offer the astronomical community a low-cost CubeSat solution for near-UV/optical/near-IR spectroscopy that enables this type of observations. 

CUBESPEC is a generic spectrograph that can be configured with minimal hardware changes to deliver both low resolution (R\,=\,100) with very large spectral coverage (200\,--\,1000\,nm), as well as high resolution (R\,=\,30\,000) over a selected wavelength range. It is built around an off-axis Cassegrain telescope and a slit spectrograph with configurable dispersion elements. CUBESPEC will use a compact attitude determination and control system for coarse pointing of the entire spacecraft, supplemented with a fine-guidance system using a fast steering mirror to center the source on the spectrograph slit and to cancel out satellite jitter. An extremely compact optical design allows us to house this instrument in a 6U CubeSat with a volume of only 10\,$\times$\,20\,$\times$\,30\,cm$^{3}$, while preserving a maximized entrance pupil of ca. 9\,$\times$\,19\,cm$^{2}$. In this contribution, we give an overview of the CUBESPEC project, discuss its most relevant science cases, and present the design of the instrument.
\end{abstract}


\section{INTRODUCTION}
\label{sec:intro}  

Over the last decade, exoplanet searches and seismic studies of the interior of stars have taken a giant leap thanks to space telescopes monitoring subtle light variations of stars.  Not hampered by the Earth's atmosphere and the interruptions of the day/night cycle of ground telescopes, the sensitive photometric cameras of the MOST, COROT and Kepler space telescopes have measured month-long series of photometric light curves, discovering hundreds of transiting exoplanets and allowing astronomers to uncover the internal structure of stars using seismic techniques.   These missions are based on rather large platforms, resulting in a long development schedule, and elevated hardware and launch costs.  Long-term photometric monitoring of bright stars from more economical nanosat-platforms is nowadays also feasible, as demonstrated in e.g.\ the BRITE project \cite{Weiss14}.  

Whereas photometric instruments measure the intensity variations integrated over a wide wavelength range, spectroscopic instruments disentangle the variations at different wavelengths. Long-term spectroscopic monitoring from space can answer a variety of exciting astrophysical questions, but is challenging to achieve from small satellite platforms. Spectrometers on larger platforms, e.g.\ the Hubble space telescope, are hugely oversubscribed and too costly to dedicate to long-term observation programs of individual targets.  

The miniaturisation trend in Space Technology now allows us to develop low-cost platforms.  Specifically, the CubeSat platform provides a cost-effective and modular standard to build, launch and operate nano-satellites (1--10\,kg) for a variety of applications. 
CubeSats are thus low-cost nano-satellites built out of standard cubic units of 10\,$\times$\,10\,$\times$\,10\,cm$^{3}$, several of which can be combined to construct larger CubeSats (e.g.\ 3U: 10\,$\times$\,10\,$\times$\,30\,cm$^{3}$ or  6U: 10\,$\times$\,20\,$\times$\,30\,cm$^{3}$). The use of a standardized form factor and off-the-shelf components lead to a relatively short development time and low cost for a CubeSat mission. Furthermore, the CubeSat standard enables access to the large and cost-effective CubeSat launch opportunity market, and use of widespread CubeSat ground stations and networks. 

\begin{figure}
\begin{center}
   \resizebox{13cm}{!}{\includegraphics{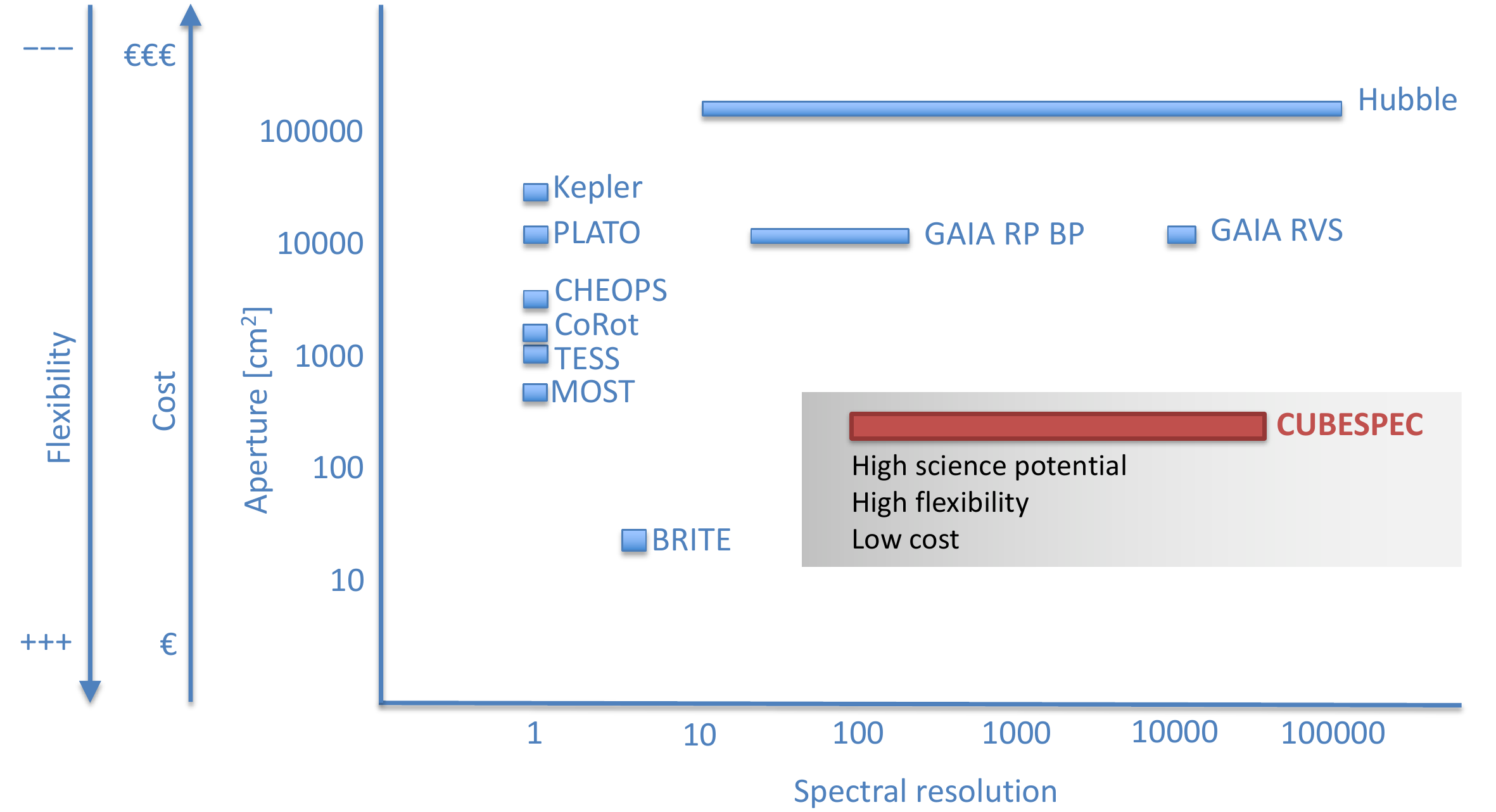}}
\end{center}
\caption 
{ \label{fig:astrosats} 
Spectral resolution and telescope collecting area of some operational, past, and planned optical space instruments. Larger aperture is also a proxy for higher cost and lower flexibility. CUBESPEC is covering a unique niche in the high science potential / high flexibility / low cost area of this diagram.}
\end{figure}

This makes the CubeSat platform an attractive option over larger spacecraft to fill a niche of dedicated astrophysical missions (Figure~\ref{fig:astrosats}). Because of their affordability, CubeSats are ideal for the specialized study of a limited sample of bright stars.  A relatively simple instrument can be used, avoiding compromises and trade-offs that are unavoidable when designing an instrument aimed at a broader range of applications.

Spectroscopy from small, low-cost space platforms can be superior to ground-based spectrometers when the measurements require:
\begin{itemize}
\item[-]  
Long-term (weeks to months) sampling of uninterrupted series of spectra;
\item[-]  
Short-term (hours) stability, not disturbed by variations in the Earth atmosphere;
\item[-] 
Wavelengths blocked or severely absorbed by the Earth atmosphere.
\end{itemize}

CUBESPEC is a compact telescope plus a high-resolution echelle spectrograph that occupies 4 units of a 6-unit (6U) CubeSat. In its current design, this instrument offers a spectral resolution of $R = \lambda/\Delta\lambda = 30\,000$ and covers the wavelength range from 350 to 580\,nm in a single exposure. The main science goal of CUBESPEC is the uninterrupted observation of massive stars in order to unravel their internal structure through the analysis of line profile variations.
However, the possible applications of CUBESPEC are much wider, targeting low and medium resolution from UV (200\,nm) to the near-IR (1000\,nm) too. The CUBESPEC spectrograph will be sufficiently generic to allow customization to specific needs (spectral resolution, wavelength range) with minimal hardware changes, such as replacing the dispersion element(s) or the detector.

\section{Science case}
\label{sec:science}  
The generic design of the  instrument and the  mission allows CUBESPEC to address multiple promising science cases. We briefly describe some of them below to illustrate the scientific value of the CUBESPEC project.

\subsection{The inner structure of massive stars}
With a mass of 8 to 300 times that of our Sun, massive stars are are key astrophysical objects that heat and enrich the interstellar medium and largely contribute to the evolution of their host-galaxy.  Despite their importance, the evolution of massive stars remains poorly constrained \cite{Langer12}. In particular, their internal structure (including the size of their convective core, the effective overshooting length, and internal mixing processes) is hard to fathom. The internal stratification is however of key influence on their lifetime, the type of end-of-life explosions that they produce and the nature of the compact objects that they leave behind.  

\begin{figure}
\begin{center}
   \resizebox{13cm}{!}{\includegraphics{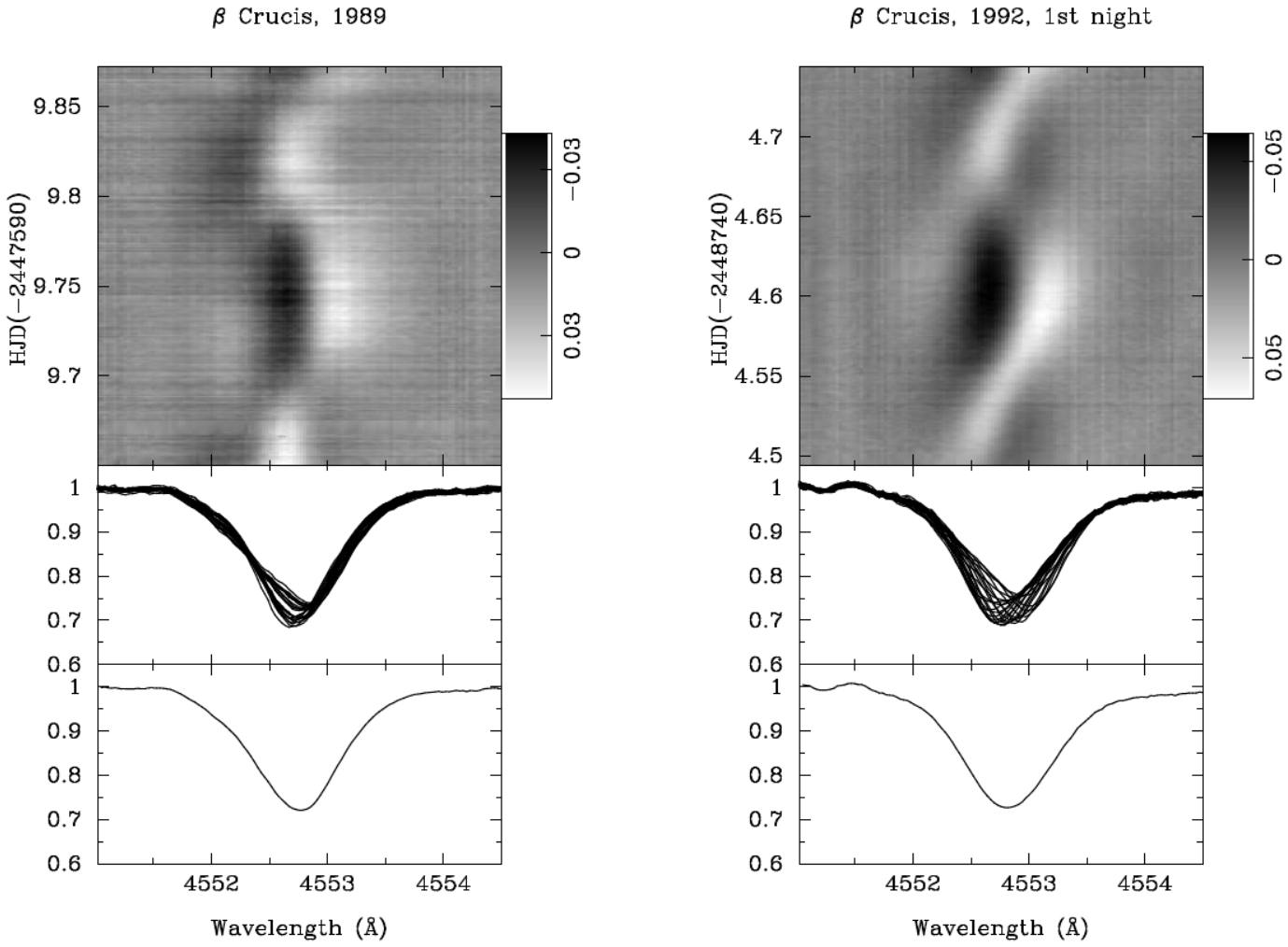}}
\end{center}
\caption 
{ \label{fig:asteroseismology} 
Grey-scale representations (top panel) of the line profile variations (of which some are shown in the
middle panel) of the Si III line at $\lambda = 4553$\,\AA\ of the $\beta$-Cep star $\beta$\,Crucis (visual magnitude of 1.25), measured in
different years. The average profile of the night is shown in the bottom panel and was subtracted from each of
the measured profiles, after which the residual flux at each wavelength pixel was assigned a grey value
according to the scale (in continuum units) shown on the right of each upper panel (white means less
absorption than for the average profile, black means deeper absorption than the average) (Figure from Ref.~\citen{Aerts98}).
}
\end{figure}

Through the study of stellar pulsations, asteroseismology has the ability to constrain the internal structure of stars with spectral type earlier than B2 (birth mass ca. 8\,$M_{sun}$), which are predicted to be either beta Cep type pulsators, slowly pulsating B stars (SPB) or a hybrid between the two \cite{Moravveji2016} (Figure~\ref{fig:asteroseismology}). Robust  identification of pulsation modes -- a pre-requisite for proper interpretation of the observations --  however requires high-quality spectroscopic time series\cite{Briquet2003}, a spectral resolution of at least $R = 30\,000$ and a signal-to-noise ratio of $\sim$100. So far this has only been pursued using ground-based telescopes and these observations suffer from time gaps that limit the disentangling of the pulsation frequencies and hence, the scientific value of the data. Indeed, these gaps  have long been the major problem in the study of time-dependent phenomena where the variability timescales are of the order of a day or shorter, as it is for g-modes SPB stars (with 1-day periods) \cite{deCat2002} and p-modes in beta-Cep pulsators (with 6 to 12 hour periods\cite{Handler2005}). The availability of uninterrupted space-based spectroscopy for even a small number of such targets would provide an enormous step forward for asteroseismology of massive stars.

\subsection{Low and intermediate mass stars}
While the previous science case requires a high spectral resolving power, the most promising customization of the CUBESPEC platform for lower mass stars is a low-resolution grating providing a  wavelength coverage from the near-ultraviolet (NUV) to the near-infrared (NIR). Such a configuration  will provide  a unique insight in phenomena that require simultaneous multi-wavelength and time-series observations. 

\begin{enumerate}
\item[-] {\bf Stellar activity}: 
 Solar-like activity in  low- and intermediate-mass stars is driven by a dynamo generating magnetic fields, which  periodically emerge through the photosphere and cause spots on the surface\cite{Schrijver2000}. During this process, magnetic fields can reconnect and cause flares and coronal mass ejections. The mechanisms underlying these cycles are far from understood, especially their connection to the stellar structure and physical processes such as rotation. The UV radiation, stellar winds, spots and mass ejections associated with activity have important, but poorly understood, effects on the evolution of planetary atmospheres and the habitability of planets.  Due to the  Earth's atmosphere, data in the UV part of the spectrum -- the wavelength range most affected by stellar activity -- are difficult to obtain and require space-based observations \cite{Wehrli2013}. 

\item[-] {\bf Fundamental stellar parameters}:  
The large wavelength coverage of the proposed low-resolution configuration is ideal to  homogeneously measure the spectral energy distribution, and  to determine the bolometric flux of stars. Combined with interferometrically measured angular diameters, this will provide a uniform calibration of the stellar effective temperature scale \cite{Boyajian2013} with unprecedented accuracy. 
While this is often neglected, the determination of the most fundamental stellar parameter, the effective temperature, is indeed in almost all cases model-dependent, and thus potentially not very accurate\cite{Ramirez2005, Ligi2015}.  In addition, multi-color information of known binaries throughout their eclipses has the potential to uniquely characterise  the limb-darkening effects \cite{Claret2008} -- which often limit the precision to which stellar parameters can be derived from binary light curves -- and  provide a stringent test for the new generation of  3D stellar atmosphere models.  
\end{enumerate}

\subsection{Exoplanets}
Almost four thousand exoplanets are known to date but  so far, only a small fraction  orbit 
 stars  bright enough to allow for a characterisation of their atmospheres. In this context, the Transiting Exoplanet Survey Satellite (TESS) \cite{Ricker15}  will be a game changer as it is expected to discover thousands of exoplanet candidates in orbits around the brightest stars in the sky. To investigate the habitability of these planets, three features will need to be  further investigated: their transmission spectra, their thermal spectra, and their phase curves. Low resolution transmission spectroscopy covering the NUV, optical and NIR can help us to shed light on the diversity of exoplanetary compositions and, in consequence, into their formation and evolution histories.

 \begin{enumerate}
\item[-] {\bf Transmission spectroscopy of exoplanets}: Transiting planets orbiting bright stars ($V < 7$), which orbit very close to their host stars (within few days periods), are currently the best targets for transmission spectroscopy with CubeSat technology. Indeed, atmospheric studies based on transit observations, and more specifically transmission spectroscopy in the NUV and visible offer an avenue for characterizing exoplanets in detail. 
Monitoring the stellar host flux over long periods of time is also important because this will teach us how stellar irradiation is absorbed, circulated, and re-emitted in exoplanet atmospheres. These are indeed important factors that control planetary evolution, climates and appearances.

\item[-] {\bf Host star monitoring: } Stellar activity has two important consequences for exoplanet atmospheric studies: one is an observational challenge, the other is astrophysical. Observationally, the spectrum of an active star changes for different levels of activity (spots), and this difference matters significantly at the level of precision that is required to probe exoplanet atmospheres ($\sim$100\,ppm/10\,nm). This variability affects the observations of exoplanets transmission spectra  and must be corrected to obtain the true planetary transmission spectrum.
The effects are however wavelength dependent so that the methods proposed to correct exoplanet spectra require long term multi-wavelength photometric monitoring (from a month to years) of host stars \cite{Pont2013, Sing2011, McCullough2014}. From an astrophysical viewpoint, how planetary atmospheres respond to energetic radiation and events from their host stars has mainly been approached theoretically and semi-empirically but lack observational constraints. CUBESPEC can be tuned to provide  the NUV and/or optical  long term multi-wavelength photometric time series  that are critically needed for estimating the flux received by these planets. These observations are crucial since visible and NUV photons drive the dissociation of molecules, produce protective layers, and trigger atmospheric escape processes, thus determining the planets ultimate habitability. 
\end{enumerate}

\section{CUBESPEC design}
\label{sec:design}  

\subsection{Spacecraft overview}

\begin{figure}
\begin{center}
   \resizebox{\hsize}{!}{\includegraphics{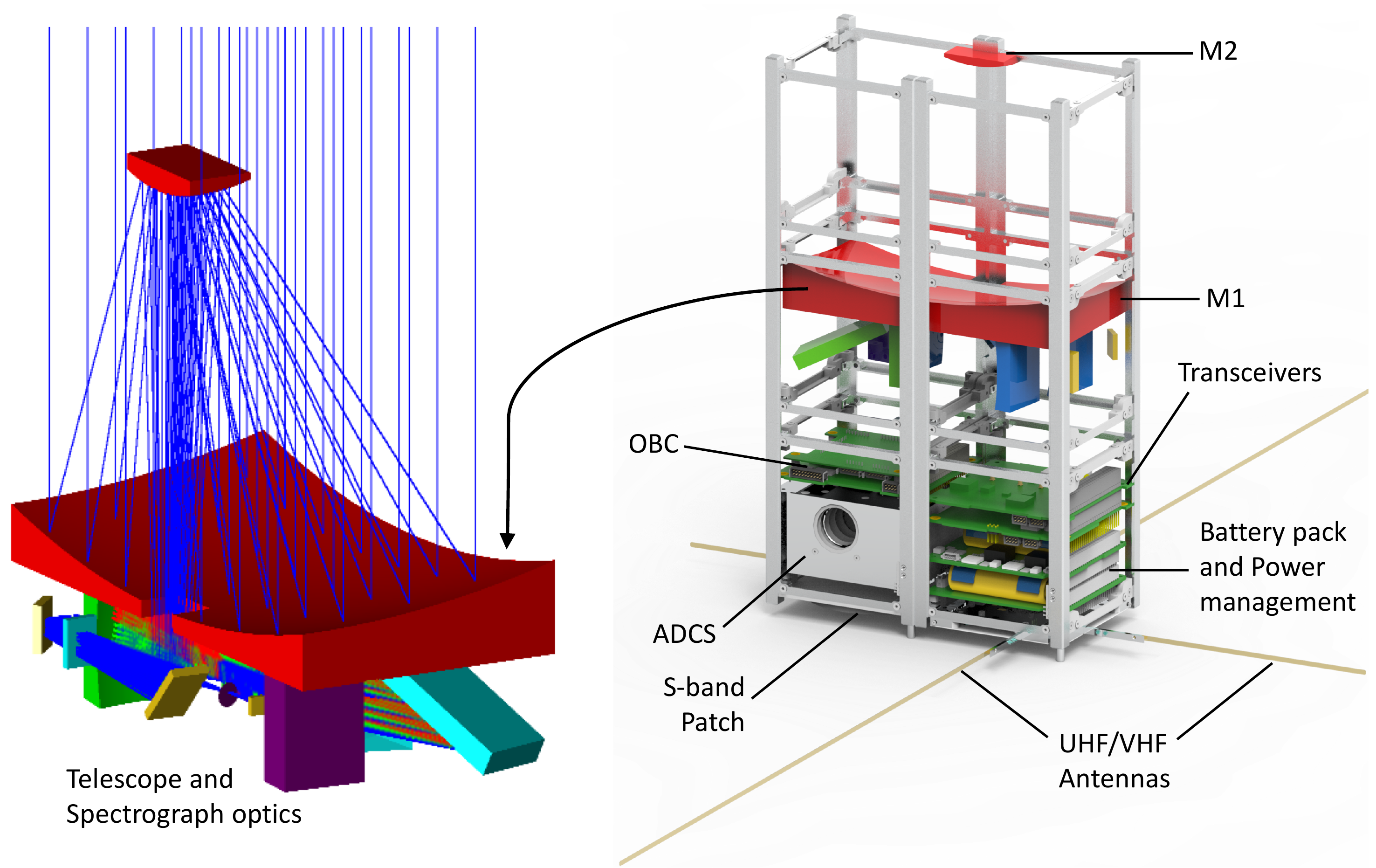}}
\end{center}
\caption 
{ \label{fig:overview} 
Left: CUBESPEC optics; the spectrograph optics are mounted on the rear surface of the telescope primary mirror (M1), the secondary mirror (M2) is located at the top of the spacecraft. Right: overview of the CUBESPEC spacecraft in a 6U CubeSat structure, the side cover and solar panels are omitted for clarity.
\vspace{6mm}}
\end{figure}

A schematic overview of the CUBESPEC spacecraft is given in Figure~\ref{fig:overview}. The top four units of the 6-unit spacecraft are occupied by the payload, consisting of a two-mirror telescope and a spectrograph. The bottom two units are reserved for the spacecraft bus.
The spacecraft bus consists of an On Board Computer (OBC) that controls all other subsystems. The compact Attitude Determination and Control System (ADCS) controls the coarse attitude of the spacecraft. A UHF/VHF communications system is used for housekeeping communication with the ground station. For science data transmission, an S-band or X-band communication will be used. With an S-band antenna, around 20 MB per day per ground station (in Belgium) can be downlinked. With an X-band antenna, even higher data rates are available. This is sufficient for downlinking the relevant parts of the observed CUBESPEC spectra. The electrical power system is sized to allow continuation of measurements, even in eclipse.

\subsection{Optical design}
The main challenges to enable CubeSat-based spectroscopy is to achieve an extremely compact state-of-the-art optical design, maximising the telescope aperture and delivering high throughput and spectral resolving power within a very limited volume and power budget. The CUBESPEC design  consists of a telescope, followed by the spectrograph folded at the rear side of the telescope primary mirror (M1). As detector, we propose to use a space-qualified CCD47-20  back-illuminated frame-transfer CCD from e2v (1k\,$\times$\,1k 13-$\mu$m pixels). 

The CUBESPEC telescope is an off-axis Cassegrain telescope, occupying $\sim$70\% of the 4U (10\,$\times$\,10\,$\times$\,20\,cm$^{3}$) payload volume. The rectangular aperture of the primary mirror (92\,$\times$\,192\,mm$^{2}$) utilizes almost the entire 2U acreage. The space constraints imply that the separation between the two telescope mirrors has to be very small (136\,mm). Together with a focal length of  1600\,mm, this leads to strong mirror curvature. As a consequence, the telescope is very sensitive to misalignment and a thermally insensitive mirror mounting is imperative. Therefore, the mirrors will be manufactured from low-CTE material and M2 will be mounted on an Invar truss.

A 45$^{\circ}$ fold mirror behind the cut-out in M1 allows us to layout the spectrograph optics on the rear surface of M1. As most of these optics consist of the same low-CTE material as the telescope mirrors, glueing them on the back of M1 results in an overall  system with excellent thermal stability. The 45$^{\circ}$ fold mirror also serves as beam steering mirror to stabilise the position of the stellar target on the slit of the spectrograph by means of a Fine Guidance Sensor (FGS, section~\ref{sec:FPP}). A beam splitter transmits a small part of the flux to the FGS, while the remainder is reflected to the spectrograph.
A three-mirror-anastigmat (TMA) collimates the beam that passes through the slit, and projects it on the dispersing element. The dispersed beam is then re-imaged on the detector after a second pass through the same TMA.

%

%
\begin{figure}
\begin{center}
   \resizebox{\hsize}{!}{\includegraphics{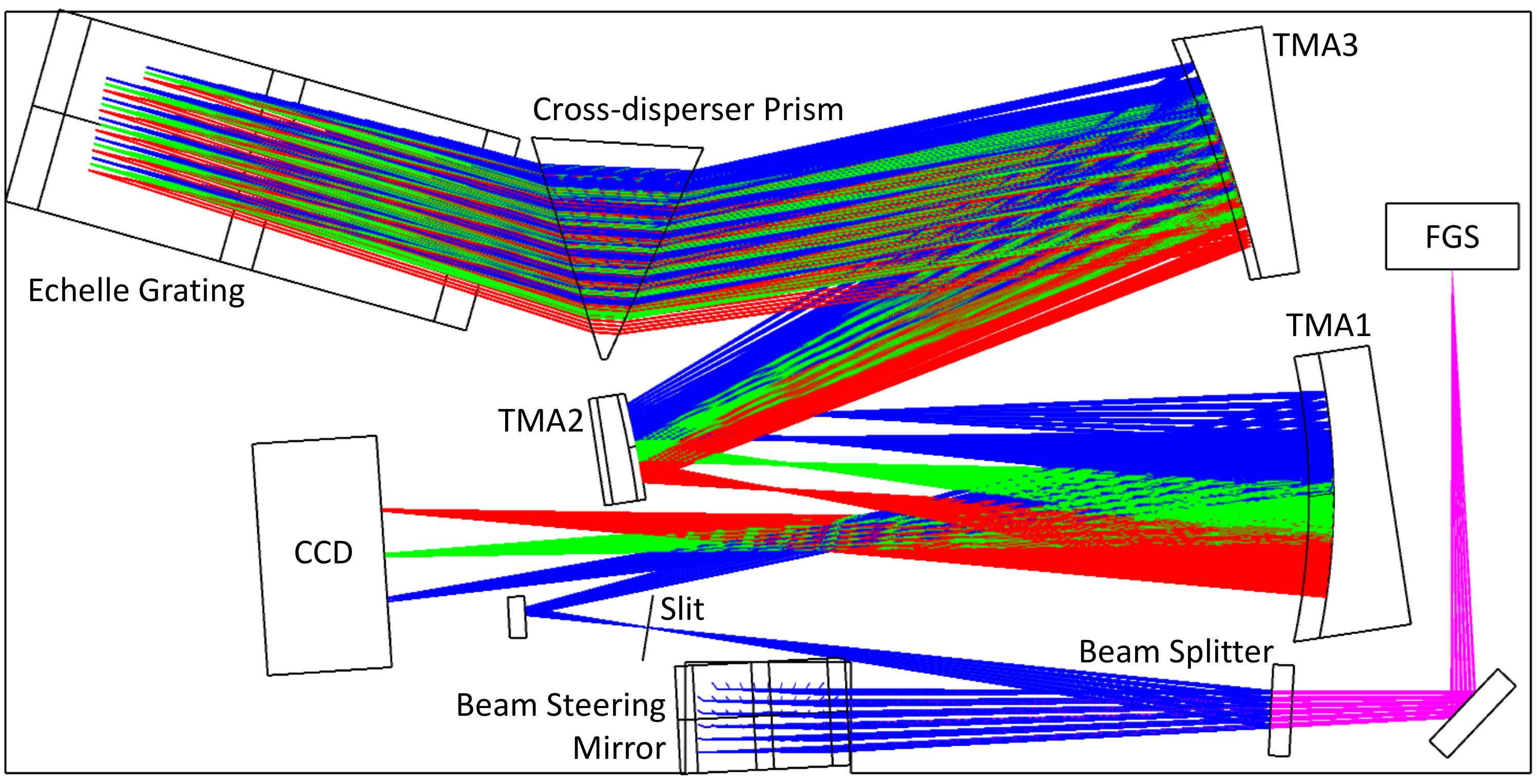}}
\end{center}
\caption
{ \label{fig:raytrace} 
Layout and raytrace of the high-resolution echelle spectrograph.}
\end{figure}

\begin{figure} 
  \begin{center}
    \resizebox{12.5cm}{!}{\includegraphics[angle=270]{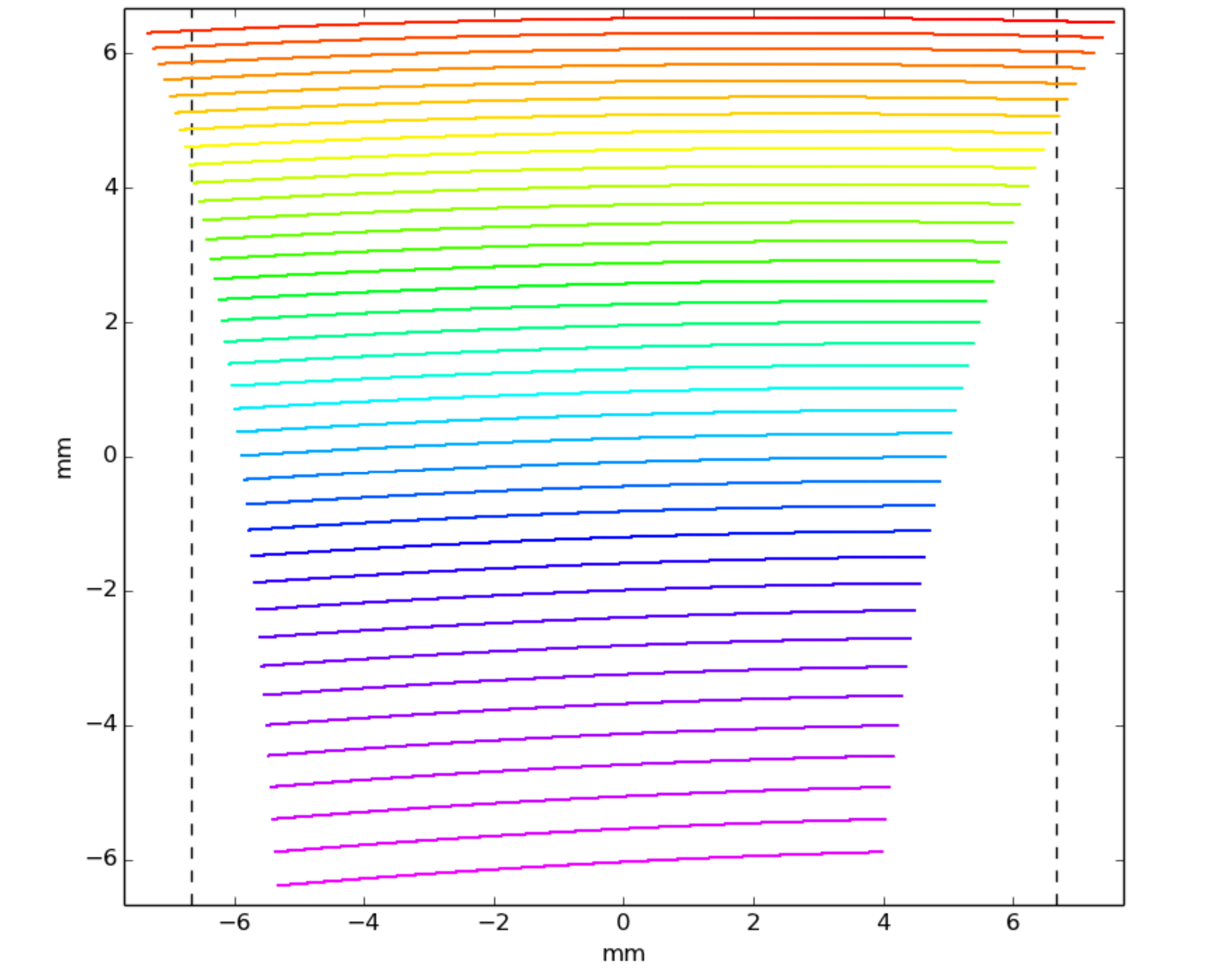}}
  \end{center}
\caption
{ \label{fig:echellogram} 
False-color simulation of how a two-dimensional echelle spectrum fits on the detector. Each line corresponds with one spectral order. The echellogram spans most of the spectral range from 350\,nm (bottom left) to 580\,nm (top right) in 37 spectral orders. The dashed vertical lines at 6.67\,mm correspond with the edges of the 1k CCD detector, showing that the reddest orders are slightly truncated.
\vspace{6mm}
}
\end{figure}

The high-resolution spectrograph uses an echelle diffraction grating (52.67\,grooves/mm, blazed at 63.5$^{\circ}$, commercially available from Richardson Gratings -- Newport), combined with a cross-dispersing prism in double pass. The cross-dispersing prism separates the overlapping diffraction orders and produces a two-dimensional spectrum that nicely matches the square detector. This allows us to combine a wide wavelength range with high spectral resolution on a compact detector. In a typical layout, the spectral resolution is $R=30\,000$ and the spectrograph observes almost the complete spectrum from 350\,nm  up to 580\,nm in 37 spectral orders. Figure \ref{fig:echellogram} shows a simulation of how such an echelle spectrum is imaged on the detector. By adjusting the rotation of the prism, we can tune the instrument for operation at shorter or longer wavelengths, or alternatively, we can select an echelle grating with finer (79\,gr/mm) or coarser (31.6\,gr/mm ) ruling for UV or near-IR operation, respectively.

In case of a low-resolution instrument, the echelle grating is replaced by a fused silica prism ($33^{\circ}$ apex angle) with high UV transmission. This prism has a mirror coating on its rear surface in order to operate in double pass. The non-linear dispersion characteristics of the fused silica make that the spectral resolution increases from $R=80$ in the near-IR to more than 1200 in the UV. At these resolution values, the 1k~detector accommodates the complete spectrum ranging from 220\,nm to 950\,nm. Variable binning can be used to obtain more homogeneous resolution values.
The above mentioned resolution values are obtained with a slit that is 28\,$\mu$m wide (slightly more than 2 CCD pixels), corresponding to a sky aperture of 3.6\,arcsec. This is substantially more than the diffraction limit of the telescope (0.67 or 1.33\,arcsec at 500\,nm) and allows us to relax the requirements on the platform pointing accuracy and stability to practical values. 

\begin{figure} [ht]
  \begin{center}
     \resizebox{\hsize}{!}{\includegraphics[angle=0]{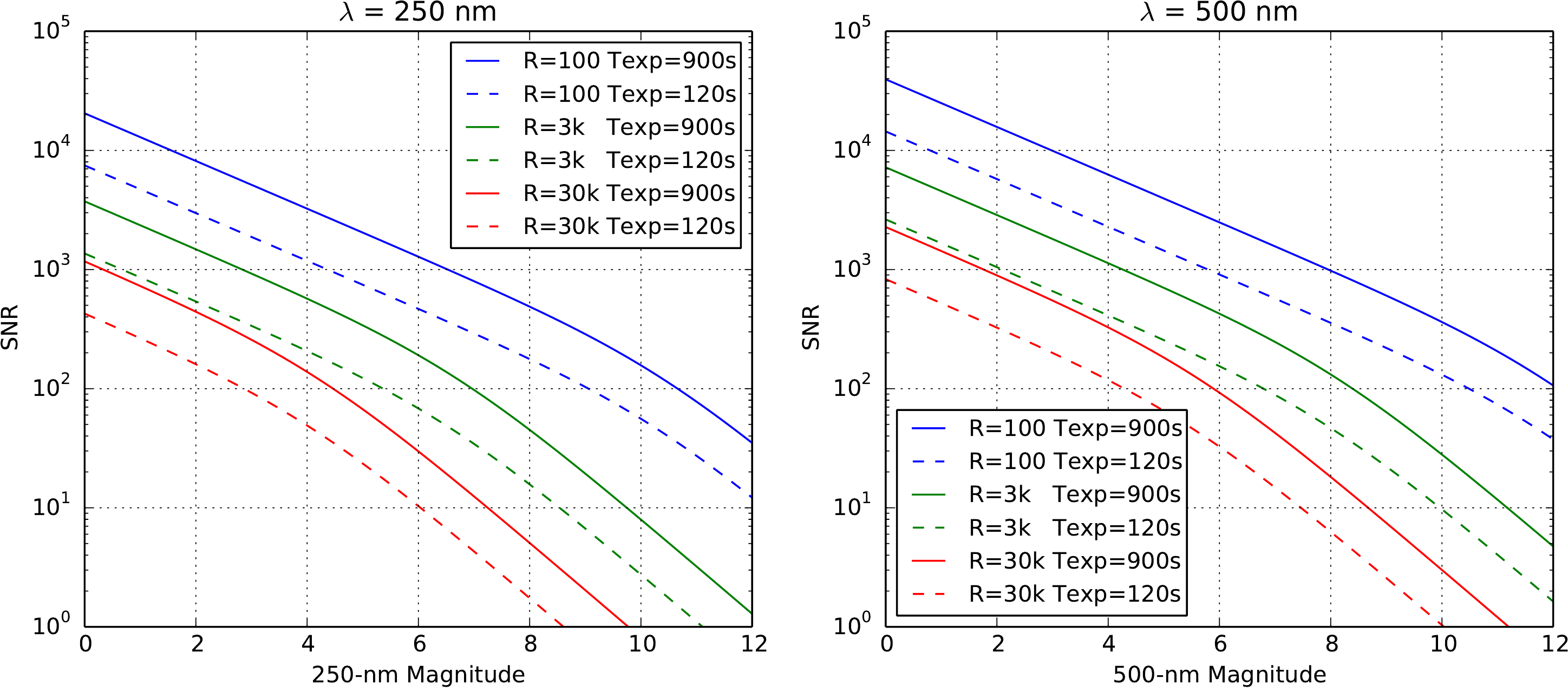}}
  \end{center}
\caption 
{ \label{fig:snr} 
Expected sensitivity (SNR per wavelength bin) of a grating ($R=3000$ and $R=30\,000$) and prism ($R=100$ spectrograph at 250\,nm (left) and 500\,nm (right) with an exposure time of 2 and 15 minutes.}
\end{figure}

In Figure~\ref{fig:snr}, we show the signal-to-noise ratio (SNR) that we expect to achieve with CUBESPEC, for a range of stellar magnitudes, both in the UV and the visible domain. We have assumed conservative estimates for the quantum efficiency of the detector and for the throughput of all optical elements. Some throughput loss due to the CubeSat's pointing jitter is also included in our calculations, assuming that the target will be correctly positioned in the spectrograph slit during 80\% of the exposure time. Simulations were carried out for a low ($R=100$),  intermediate ($R=3000$) and  high ($R=30\,000$) resolution case, and for typical exposure times of 2 and 15 minutes. 

\subsection{Attitude determination and control}
A critical issue for almost every space-borne astronomical instrument is the pointing accuracy and stability. This is especially true for the CUBESPEC application, having a limited field of view and requiring accurate positioning of the target on the spectrograph slit. Moreover, small platforms such as CubeSats are inherently more unstable than large satellites. Therefore, we have to impose strong requirements on the attitude determination and control system (ADCS). The ADCS is responsible for despinning the spacecraft after launch and for performing attitude manoeuvres. 

\begin{figure}
  \begin{center}
     \resizebox{10cm}{!}{\includegraphics[trim={40mm 0mm 20mm 0mm},clip]{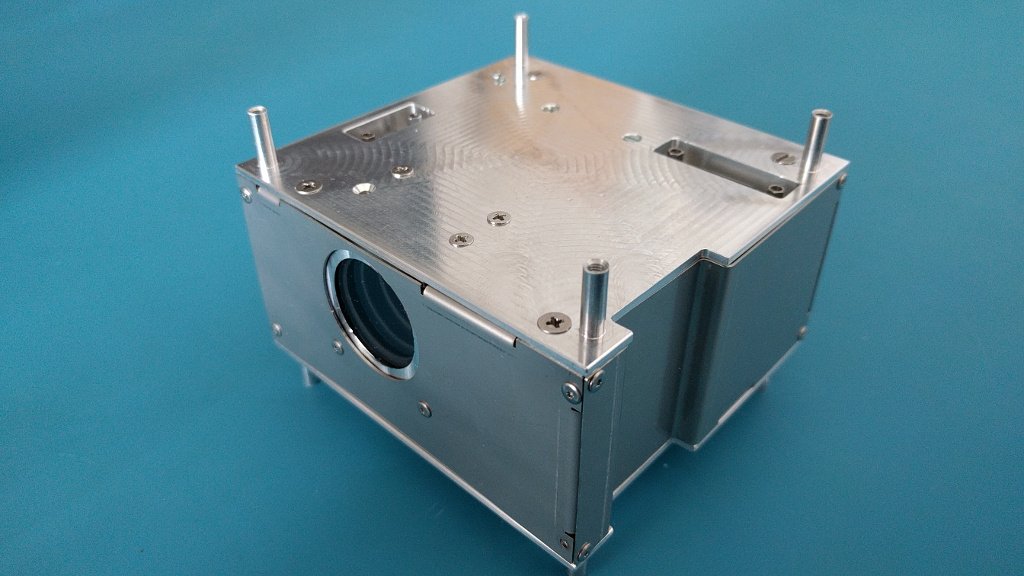}}
  \end{center}
\caption 
{ \label{fig:adcs} 
Picture of the 0.5U KUL ADCS, containing three reaction wheels, three magnetorquers and a star tracker (circular aperture in left-front side).}
\end{figure}

CUBESPEC relies on the KUL ADCS, developed at KU Leuven \cite{Delabie17, Delabie18}. This is a complete but nevertheless very compact ADCS, occupying only half of a CubeSat unit (5\,$\times$\,10\,$\times$\,10\,cm$^{3}$, see picture in Figure~\ref{fig:adcs}). It uses three reaction wheels to have agile and accurate control and three magnetorquers for reaction wheel desaturization and despinning of the satellite. The attitude determination is done using a set of sensors, of which the star tracker \cite{Delabie16} is the most accurate. 
The ADCS can determine the attitude with an accuracy in the range of a few arc seconds thanks to this star tracker. The 3$\sigma$ absolute pointing error has been analysed in the ESA PEET software \cite{Ott11, Ott14} to be below 0.1$^{\circ}$ around each axis. This is quite close to the lower limit of around 150\,arcsec for 3U-CubeSat ADCS performance as analysed in Ref.~\citenum{Vandersteen16}.

In order to reach the stringent pointing requirements of the CUBESPEC mission, the ADCS will be supplemented with a fine pointing system using a fast steering mirror. The ADCS is used for coarse pointing of the spacecraft and needs to ensure that the attitude remains close enough (better than 0.1$^{\circ}$) to the desired attitude such that the actuators of the fast steering mirror can operate within their range.

\subsection{Fine pointing platform}
\label{sec:FPP}
\begin{figure} [b]
  \begin{center}
     \resizebox{\hsize}{!}{\includegraphics[angle=0]{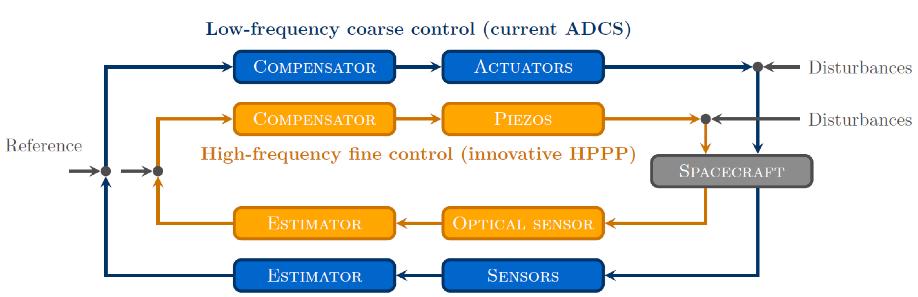}}
  \end{center}
\caption 
{ \label{fig:hppp} 
Overview of the CUBESPEC pointing control.}
\end{figure}

\begin{figure}
  \begin{center}
     \resizebox{\hsize}{!}{\includegraphics[angle=0]{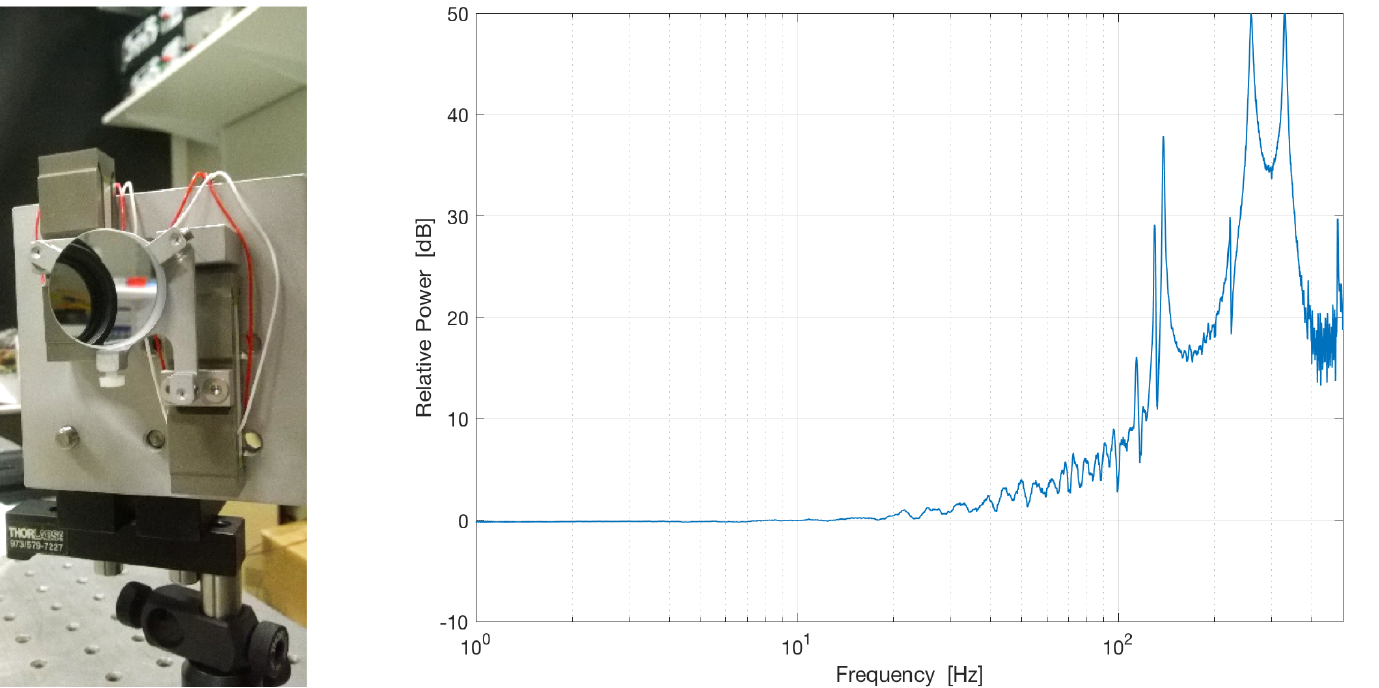}}
  \end{center}
\caption 
{ \label{fig:steeringmirror} 
Left: Picture of a prototype of the beam-steering mirror on two amplified piezo actuators ; Right: measured frequency response of the beam-steering mirror. 
\vspace{4mm}
}
\end{figure}

To avoid substantial light loss at the spectrograph slit, the total jitter should not be larger than the slit width (3.6\,arcsec), which is well below the pointing stability of the platform. To correct for the residual pointing errors and the instability of the platform ADCS, a Fine Guidance Sensor (FGS) in the spectrometer optics and a beam-steering mechanism in the fold mirror are included. The FGS pick-off is a beam splitter (partially reflective or a dichroic) that reflects part of the beam to the FGS while most of the light is transmitted to the spectrograph. The steering mirror will be mounted on three piezo-electric actuators that provide arcsec positioning precision. An accurate control loop, as illustrated in Figure~\ref{fig:hppp}, takes care of stabilizing the star image on the FGS and, at the same time, locking this image in the spectrograph slit. Apart of tip/tilt movement, the three  piezos also provide some piston that will be used for focus corrections.
We already built  a prototype design of the beam-steering mirror (Figure~\ref{fig:steeringmirror}) using two amplified piezo actuators and one fixed pivot, to perform initial tests in order to verify the feasibility of the fine pointing approach. In the right part of Figure~\ref{fig:steeringmirror} we show the measured frequency response of this assembly, with a first Eigen-frequency above 100\,Hz, sufficiently more than the bandwidth of the fine pointing control loop.

\begin{figure}
  \begin{center}
     \resizebox{\hsize}{!}{\includegraphics[angle=0]{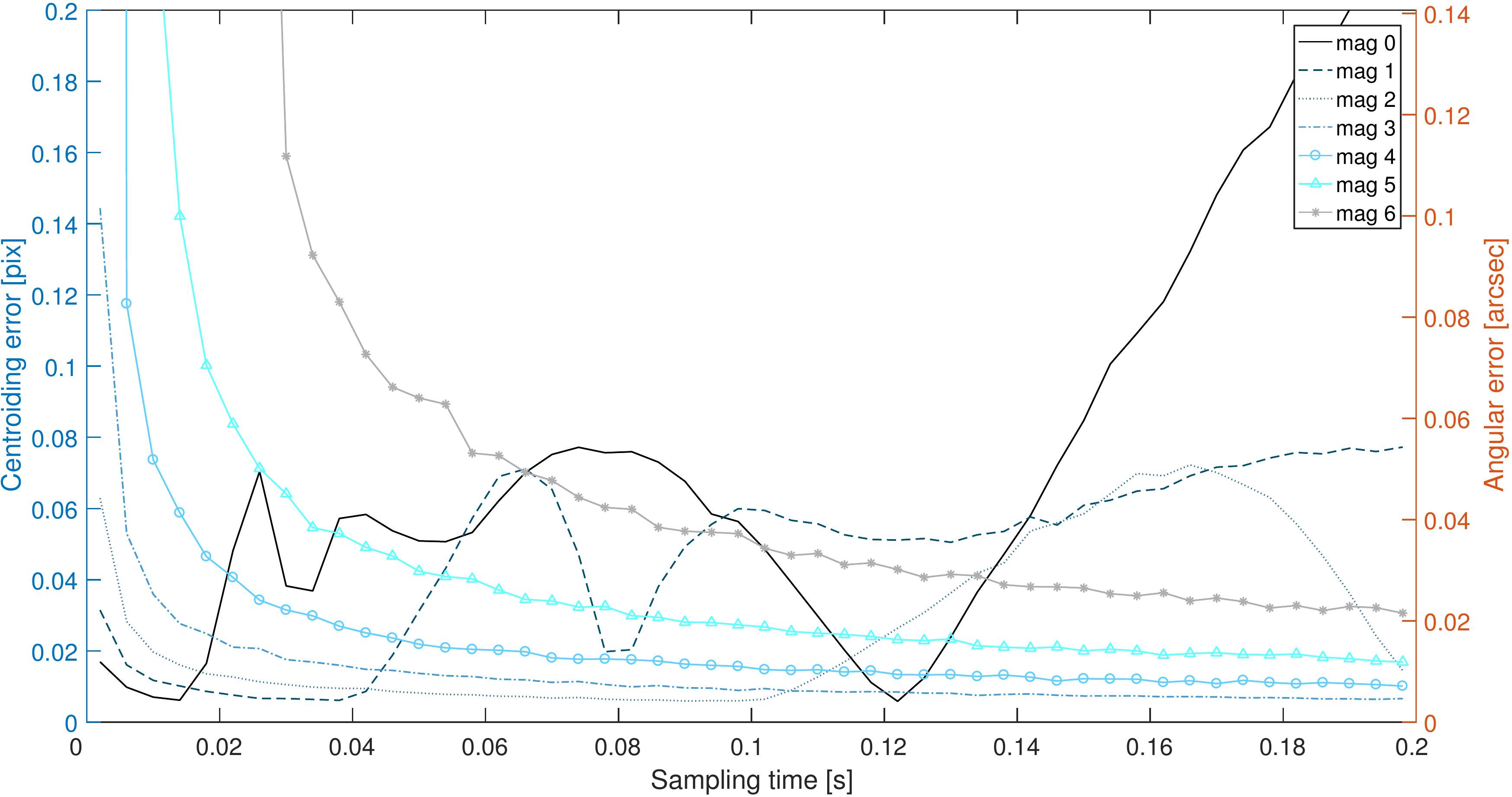}}
  \end{center}
\caption 
{ \label{fig:fgs} 
FGS accuracy in function of sampling time for different magnitude sources.
\vspace{5mm}
}
\end{figure}

The FGS uses a CMV4000 cmos image sensor from CMOSIS (2k\,$\times$\,2k, 5.5\,$\mu$m or 1.4\,arcsec pixels) to measure the position of the observed star and calculate the deviations due to attitude disturbances. Based on this information, the required action of the steerable mirror can be determined.
It is important to provide attitude information at a high rate in order to control the steering mirror. While the piezo-actuated mirror has a sufficiently large bandwidth, the FGS flux levels required for accurate centroiding will be the main bottleneck for the speed of the fine pointing control loop. 
We simulated the centroiding accuracy on generated star images, assuming only 5\% of the light captured by the telescope is transmitted to the FGS and including realistic noise values. The Gaussian Grid centroiding algorithm developed at KU Leuven \cite{Delabie16b} was used to determine the star centroid. The RMS values of the centroiding error in function of sampling time are given for a range of target magnitudes in Figure~\ref{fig:fgs}. A good trade-off can be found when the FGS is sampled at 60\,Hz. At that point, stars up to 5$^{th}$ magnitude can be centroided with an accuracy close to 0.1\,arcsec and we should be able to correct the pointing at a rate of ca. 10\,Hz. In case the FGS is allowed to use 10\% of the total flux, we can get acceptable centroiding up to 6$^{th}$ magnitude or increase the correction speed to 20\,Hz.
We analysed the 3$\sigma$ absolute pointing error of the two-stage approach (ADCS + fine pointing platform)  with these settings in the ESA PEET software and found it to be smaller than 2\,arcsec around each axis. This is well below the 3.6-arcsec spectrograph slit width and satisfies the CUBESPEC pointing requirement. Now we are planning test bench simulations to verify this absolute pointing performance.

\section{Project status and planning}
\label{sec:status}  

The CUBESPEC mission and payload concept so far has been studied from internal resources at KU Leuven.  We are currently  in the process of planning and proposing  a technology development project with the European Space Agency, aiming at a phase-A study of 1 year, and phases B, C \& D extending over 2 years. Therein we aim at an in-orbit demonstration on a time-scale of 3 years.  This project will be a collaboration between academic institutions and industrial partners.

The long-term goal of the project is to provide a generic low-cost, recurrent off-the-shelve solution to deliver spectroscopic monitoring of point sources from space.  Customer's requirements on spectral resolution and spectral coverage shall require only minimal hardware changes or software configuration, facilitating a quick order-to-launch turnaround at a minimised cost. This way, CUBESPEC will make customised space-based spectroscopy  affordable for a large part of the astronomical community.

\acknowledgments 
 The authors thank Timothy R. White and Carolina Von Essen for valuable input to the CUBESPEC science case.
 Wim De Munter is an SB PhD fellow at Research Foundation – Flanders (FWO).
Funding for the Stellar Astrophysics Centre is provided by The Danish National Research Foundation (Grant agreement no.: DNRF106). We thank the Aarhus University Research Foundation (AUFF) for their financial  support. 

\bibliography{report} 
\bibliographystyle{spiebib} 

\end{document}